\documentclass[twocolumn,superscriptaddress,showpacs,preprintnumbers,amsmath,amssymb,prc]{revtex4}

\usepackage{graphicx,color}
\usepackage{dcolumn}
\usepackage{bm}
\usepackage{mathrsfs}
\usepackage{amsmath}

\begin{document}

\title{Functional Renormalization Group and Kohn-Sham scheme in Density Functional Theory}

\author{Haozhao Liang}\email{haozhao.liang@riken.jp}
 \affiliation{Nishina Center, RIKEN, Wako 351-0198, Japan}
 \affiliation{Department of Physics, Graduate School of Science, The University of Tokyo, Tokyo 113-0033, Japan}

\author{Yifei Niu}\email{yifei.niu@eli-np.ro}
 \affiliation{ELI-NP, ``Horia Hulubei'' National Institute for Physics and Nuclear Engineering, RO-077125 Bucharest-Magurele, Romania}

\author{Tetsuo Hatsuda}\email{thatsuda@riken.jp}
 \affiliation{iTHEMS Program and iTHES Research Group, RIKEN, Wako 351-0198, Japan}
 \affiliation{Nishina Center, RIKEN, Wako 351-0198, Japan}

\date{\today}

\begin{abstract}
  Deriving accurate energy density functional is one of the central problems in condensed matter physics, nuclear physics, and quantum chemistry.
  We propose a novel method to deduce the energy density functional by combining the idea of the functional renormalization group and the Kohn-Sham scheme in density functional theory.
  The key idea is to solve the renormalization group flow for the effective action decomposed into the mean-field part and the correlation part.
  Also, we propose a simple practical method to quantify the uncertainty associated with the truncation of the correlation part.
  By taking the $\varphi^4$ theory in zero dimension as a benchmark, we demonstrate that our method shows extremely fast convergence to the exact result even for the highly strong coupling regime.
\end{abstract}


\preprint{RIKEN-QHP-332}

\maketitle


\textit{Introduction.}---The density functional theory (DFT) \cite{Hohenberg1964Phys.Rev.136_B864--B871, Kohn1965Phys.Rev.140_A1133--A1138} is a successful approach to reduce quantum many-body problem to one-body problem with the local density distribution $\rho(\mathbf{x})$.
Due to its high accuracy with relatively low computational cost, DFT has great success in various fields including condensed matter physics, nuclear physics, and quantum chemistry.
According to the Hohenberg-Kohn (HK) theorem \cite{Hohenberg1964Phys.Rev.136_B864--B871}, there exits an energy density functional of $\rho(\mathbf{x})$ as $E_{U}[\rho] =F_{\rm HK}[\rho] + \int \mathrm{d}^3\mathbf{x}\, U(\mathbf{x}) \rho(\mathbf{x})$, where the universal functional $F_{\rm HK}[\rho]$ is independent of the external potential $U(\mathbf{x})$.
The ground-state energy of the system corresponds to a global minimum of $E_{U}[\rho]$.
In DFT, deriving $F_{\rm HK}[\rho]$ in a systematic and controllable way is the most important issue, see, e.g., the overviews \cite{Kohn1999Rev.Mod.Phys.71_1253--1266, Kryachko2014Phys.Rep.544_123--239, Zangwill2015Phys.Today68_34--39, Jones2015Rev.Mod.Phys.87_897--923}, as well as recent topical reviews in condensed matter physics \cite{Kutzelnigg2006J.Mol.Struct.768_163--173, Berland2015Rep.Prog.Phys.78_066501}, nuclear physics \cite{Drut2010Prog.Part.Nucl.Phys.64_120--168, Nakatsukasa2016Rev.Mod.Phys.88_045004}, and quantum chemistry \cite{Cohen2012Chem.Rev.112_289--320, Himmetoglu2014Int.J.QuantumChem.114_14--49, Sperger2015Chem.Rev.115_9532--9586}.
Also, the theoretical error estimates or uncertainty quantification is a key issue in modern DFT applications \cite{Erler2012Nature486_509--512, Dobaczewski2014J.Phys.G41_074001, McDonnell2015Phys.Rev.Lett.114_122501, Nazarewicz2016J.Phys.G43_044002}.

Another successful approach to quantum many-body problem is the functional renormalization group (FRG) \cite{Wetterich1993Phys.Lett.B301_90--94}:
It is based on the one-parameter flow equation which leads to the quantum effective action at the end of the flow, see, e.g., the review \cite{Metzner2012Rev.Mod.Phys.84_299--352}.
A close connection  between the effective action $\Gamma[\rho]$ in FRG and the universal functional $F_{\rm HK}[\rho]$ in DFT has been established on the basis of the two-particle point-irreducible (2PPI) scheme, which we
call 2PPI-FRG, by Polonyi, Sailer, and Schwenk \cite{Polonyi2002Phys.Rev.B66_155113, Schwenk2004_273-282}, so that FRG provides a practical way to construct $F_{\rm HK}[\rho]$.
The 2PPI-FRG was further developed in Refs.~\cite{Braun2012J.Phys.G39_033001, Kemler2013J.Phys.G40_085105, Kemler2017J.Phys.G44_015101} with the case studies including the zero-dimensional (0-D) $\varphi^4$ theory, (0+1)-D anharmonic oscillator, and (1+1)-D Alexandrou-Negele nuclei.
See also Ref.~\cite{Rentrop2015J.Phys.A48_145002} for a comparative study.
Although the 2PPI-FRG is a systematic formalism, the resultant accuracy in these case studies was found to be not so satisfactory: Up to the next-to-leading order, the ground-state energies of (1+1)-D nuclei missed by about $30\%$ comparing to the Monte Carlo results \cite{Kemler2017J.Phys.G44_015101}.
Even for the simplest 0-D model \cite{Kemler2013J.Phys.G40_085105}, the ground-state energy still missed by about $2\%$ with the sixth-order calculation for intermediate coupling strength.
Note that the sixth-order calculations are almost infeasible for actual (3+1)-D problems, and even if it is achieved, the $2\%$-accuracy would not be good enough for practical applications of nuclear binding energies, not to mention the chemical accuracy.

The purpose of this Letter is twofold: First of all, we propose a novel optimization method of FRG in analogy with the Kohn-Sham (KS) scheme in DFT,  which we call KS-FRG.
The convergence of the energy density functional in KS-FRG is shown to be much faster than the un-optimized scheme.
Secondly, we propose a method to estimate the truncation uncertainty in the KS-FRG.
By taking the 0-D  $\varphi^4$ theory as an example, we demonstrate explicitly that these methods work well in practice.


\textit{Formalism.}---Let us consider a general non-relativistic system with a two-body interaction $V(\mathbf{x}_1, \mathbf{x}_2)$.
The bare action $S$ in the Euclidean space reads
\begin{align}\label{Eq:action}
  S[U,V] =&~ \int \psi^\dag(x)\left(\partial_\tau+K+U(\mathbf{x}) - \mu \right)\psi(x) \nonumber\\
&+{1\over 2}\iint \psi^\dag(x_1)\psi^\dag(x_2) V(\mathbf{x}_1, \mathbf{x}_2) \psi(x_2)\psi(x_1)\,,
\end{align}
with $x = (\tau,\mathbf{x})$, $\int = \int_0^{\beta} \mathrm{d}\tau \int \mathrm{d}^d\mathbf{x}$, $d$ the space dimension, $\beta$ the inverse temperature, $\mu$ the chemical potential, and $K=-\nabla^2/(2M)$.
The external potential $U(\mathbf{x})$ vanishes for self-bound systems such as atomic nuclei, while it represents  physical harmonic trap for ultracold atoms.

The generating functional of connected Green's functions is defined by
\begin{equation}
  e^{W[J]}= \int \mathfrak{D} (\psi^\dag \psi)\, \exp\{ -S[U,V] + \int J(x) \psi^\dag(x) \psi(x) \}\,,
\end{equation}
where $J(x)$ is a local external source.
The functional derivative of $W[J]$ with respect to $J$ is nothing but the local density
\begin{equation}
  {\rho}(x) = \langle \psi^\dag(x) \psi(x) \rangle = \frac{\delta W [J]}{\delta J(x)}\,.
\end{equation}
The 2PPI effective action is then defined as the Legendre transform,
\begin{equation}
  \Gamma[\rho;U,V] = -W[J] + \int J(x) \rho(x)\,,
\end{equation}
and the energy density functional at zero temperature is obtained by
\begin{equation}
  E[\rho] = \lim_{\beta\rightarrow\infty}\frac{\Gamma[\rho]}{\beta} \,.
\end{equation}

In the 2PPI-FRG formalism \cite{Polonyi2002Phys.Rev.B66_155113, Schwenk2004_273-282}, a flow parameter $\lambda\in[0,1]$ is introduced to replace $V$ by $\lambda V$ and $U$ by a {\em given} regulator function $U_{\lambda}$ with the boundary condition $U_{\lambda=1}=U$.
Then the $\lambda$-dependent 2PPI effective action is defined by $\Gamma_{\lambda}[\rho] \equiv \Gamma[\rho; U_{\lambda}, \lambda V]$ whose renormalization group flow reads \cite{Schwenk2004_273-282},
\begin{equation}\label{Eq:FRG}
    \partial_\lambda \Gamma_\lambda[\rho] =
    \rho \cdot \partial_\lambda U_\lambda +
    \frac{1}{2} \rho \cdot V \cdot \rho + \frac{1}{2} \mbox{Tr} \left\{ V \cdot \left(\Gamma^{(2)}_\lambda[\rho] \right)^{-1} \right\}\,.
\end{equation}
Here the dots and trace imply $X \cdot Y = \int X(x) Y(x)$, $X \cdot A \cdot Y = \iint X(x) A(x,y) Y(y)$, and $\mbox{Tr}\{A \cdot B\} = \iint A(x,y) B(y,x)$.
The $n$-point vertex functions are obtained by
\begin{equation}
  \Gamma^{(n)}_{\lambda; x_1, \ldots, x_n}[\rho] = \frac{{\delta^n \Gamma_{\lambda}[\rho]}}{ \delta \rho(x_1) \ldots \delta \rho(x_n)}\,.
\end{equation}

The ground-state density for a fixed $\lambda$ denoted by $\bar{\rho}_{\lambda}$ is a solution of
\begin{equation}
  \left. \frac{\delta \Gamma_{\lambda}[\rho]}{\delta \rho(x)} \right|_{\rho=\bar{\rho}_{\lambda}} = 0\,,
\end{equation}
so that the effective action $\Gamma_{\lambda}[\rho]$ can be expanded around $\bar{\rho}_{\lambda}$ as
\begin{align}
  \Gamma_{\lambda} [\rho]
  &= \Gamma_{\lambda}^{(0)}[\bar{\rho}_{\lambda}] + \frac{1}{2} \iint \Gamma^{(2)}_{\lambda; x_1,x_2}[\bar{\rho}_{\lambda}](\rho-\bar{\rho}_{\lambda})_{x_1} (\rho-\bar{\rho}_{\lambda})_{x_2} \nonumber \\
  &\quad+ \cdots \nonumber\\
  &\equiv \bar{\Gamma}_{\lambda}^{(0)} + \sum_{n=2}^\infty  \frac{1}{n!} \int \bar{\Gamma}_\lambda^{(n)} \cdot (\rho - \bar{\rho}_{\lambda})^n\,,
\end{align}
where $\bar{\Gamma}^{(n)}_{\lambda } \equiv {\Gamma}^{(n)}_{\lambda } [\bar{\rho}_{\lambda}]$.
This power series expansion together with the flow equation~(\ref{Eq:FRG}) leads to an infinite hierarchy of coupled integro-differential equations for $\bar\Gamma_{\lambda}^{(n)}$ and $\bar{\rho}_{\lambda}$.
As shown in some case studies, however, such a ``naive'' expansion converges rather slowly to the exact results \cite{Kemler2013J.Phys.G40_085105, Kemler2017J.Phys.G44_015101}.

Here we propose the KS-FRG which is a novel optimization theory of FRG with faster convergence under the same spirit with the KS scheme in DFT \cite{Kohn1965Phys.Rev.140_A1133--A1138}.
The basic idea is to introduce an effective action for a hypothetical non-interacting system with a mean-field KS potential $U_{{\rm KS}, \lambda}(\mathbf{x})$ and to split the total effective action into the mean-field part $\Gamma_{{\rm KS},\lambda}$ and the correlation part $\gamma_{\lambda}$,
\begin{eqnarray} \label{eq:gamma-def}
    \Gamma_{\lambda}[\rho] = \Gamma_{{\rm KS},\lambda}[\rho] + \gamma_{\lambda}[\rho]\,,
\end{eqnarray}
with $\Gamma_{{\rm KS},\lambda}[\rho] \equiv \Gamma[\rho; U_{{\rm KS},\lambda}, 0]$.
These two terms are determined simultaneously by solving the FRG flow equation together with the KS equation.

Explicit form of the self-consistent equation to obtain  $\Gamma_{{\rm KS},\lambda}[\rho]$
through $U_{{\rm KS}, \lambda}$ is
\begin{eqnarray}\label{Eq:KS-eq}
    \left. \frac{\delta \Gamma_{{\rm KS},\lambda}[\rho]}{\delta \rho(x)} \right|_{\rho=\bar{\rho}_{\lambda}} = 0\,.
\end{eqnarray}
This implies that $\bar{\rho}_{\lambda}$ is a common stationary point for both $\Gamma_{{\rm KS},\lambda}[\rho]$ and  $\Gamma_{\lambda}[\rho]$.
Equation~(\ref{Eq:KS-eq}) is equivalent with the standard KS equation \cite{Kohn1965Phys.Rev.140_A1133--A1138, Drut2010Prog.Part.Nucl.Phys.64_120--168} written in terms of the single-particle wave functions, since it is nothing more than the one-body problem with $V=0$.
The flow equation for the correlation part is obtained from Eqs.~(\ref{Eq:FRG})--(\ref{Eq:KS-eq}) as
\begin{align}\label{Eq:FRG2}
  \partial_\lambda   \gamma_{\lambda} [\rho]
  =&~
  \rho \cdot \left( \partial_\lambda U_\lambda +  \bar{\Gamma}^{(2)}_{{\rm KS},\lambda } \cdot \partial_{\lambda} \bar{\rho}_{\lambda}  \right)  + \frac{1}{2}\rho\cdot V\cdot\rho   \nonumber \\
  & +  \frac{1}{2} \mbox{Tr}\left\{ V \cdot   \left( \Gamma^{(2)}_{{\rm KS},\lambda } [\rho] +  \gamma^{(2)}_{\lambda} [\rho]    \right)^{-1} \right\}\,.
\end{align}
Here we have used the following chain rule,
\begin{equation}
  \partial_{\lambda} \Gamma_{{\rm KS},\lambda} =
 \frac{\delta \Gamma_{{\rm KS},\lambda}}{\delta U_{{\rm KS},\lambda}}\cdot
 \frac{\delta U_{{\rm KS},\lambda}}{\delta \bar{\rho}_{\lambda}}\cdot \partial_{\lambda} \bar{\rho}_{\lambda}
 = - \rho \cdot  \bar{\Gamma}^{(2)}_{{\rm KS},\lambda }  \cdot \partial_{\lambda} \bar{\rho}_{\lambda}\,.
\end{equation}
As seen from the first term in the right-hand side, the effective one-body term proportional to $\rho$ is properly separated out.
Note also that the choice $U_{{\rm KS}, \lambda=0} = U_{\lambda=0}$ leads to the initial condition $\gamma_{\lambda=0}[{\rho}]=0$.

Equations~(\ref{eq:gamma-def}), (\ref{Eq:KS-eq}), and (\ref{Eq:FRG2}) are the  master equations in KS-FRG.
To solve them in practice, we expand the correlation part $\gamma_{\lambda}[\rho]$ around $\bar{\rho}_{\lambda}$,
\begin{equation}\label{Eq:optimize}
   \gamma_{\lambda}[\rho] =  \bar{\gamma}_\lambda^{(0)}
   + \sum_{n=2}^\infty  \frac{1}{n!}  \int \bar{\gamma}_\lambda^{(n)} \cdot
   (\rho-\bar{\rho}_{\lambda})^n\,.
\end{equation}
On the other hand, we do not introduce the expansion for the mean-field part in Eq.~(\ref{eq:gamma-def}).
This is in contrast to the case of 2PPI-FRG where the whole $\Gamma_{\lambda}[\rho]$ is expanded as a power series.

\begin{table*}
  \centering
  \caption{Ground-state densities and energies for weak-, intermediate-, and strong-coupling cases.
  The 1st-, 2nd-, and 3rd-order results of KS-FRG with theoretical uncertainties are compared with the exact values up to appropriate digit.}
  \label{Table1}
\begin{ruledtabular}
\begin{tabular}{rlllllllll}
    && \multicolumn{2}{c}{$(\omega,v)=(1,0.01)$}
    && \multicolumn{2}{c}{$(\omega,v)=(1,1)$}
    && \multicolumn{2}{c}{$(\omega,v)=(1,100)$} \\
    && \multicolumn{1}{c}{$\rho_{\rm gs}$} & \multicolumn{1}{c}{$E_{\rm gs}$}
    && \multicolumn{1}{c}{$\rho_{\rm gs}$} & \multicolumn{1}{c}{$E_{\rm gs}$}
    && \multicolumn{1}{c}{$\rho_{\rm gs}$} & \multicolumn{1}{c}{$E_{\rm gs}$} \\ \hline
  1st-order && $ {{0.995065}}29(^{+27}_{-77})$ & $ {{0.0012417}}85(^{+55}_{-19})$
            && $ {{0.7}}45(^{+18}_{-23})$ & $ {{0.08}}50(^{+11}_{-10})$
            && $ {{0.}}12(^{+13}_{-4})$ & $ {{0.}}64(^{+32}_{-18} )$ \\
  2nd-order && $ {{0.995065}}30(^{+8}_{-20})$ & $ {{0.001241778}}80(^{+31}_{-81})$
            && $ {{0.75}}08(^{+14}_{-11})$ & $ {{0.084}}56(^{+36}_{-45})$
            && $ {{0.1}}42(^{+29}_{-9} )$ & $ {{0.7}}47(^{+44}_{-17})$ \\
  3rd-order && $ {{0.99506532}}85(^{+21}_{-8})$ & $ {{0.0012417789}}44(^{+16}_{-44})$
            && $ {{0.750}}52(^{+15}_{-15})$ & $ {{0.0845}}29(^{+67}_{-68})$
            && $ {{0.14}}82(^{+58}_{-65})$ & $ {{0.75}}97(^{+52}_{-62})$ \\ \hline
  exact     && $ {{0.99506532}}82$ & $ {{0.0012417789}}51$
            && $ {{0.750}}51$      & $ {{0.0845}}57$
            && $ {{0.15}}04$        & $ {{0.75}}97$ \\
\end{tabular}
\end{ruledtabular}
\end{table*}

By expanding both sides of Eq.~(\ref{Eq:FRG2}) in terms of a dimensionless power counting parameter $(\rho-\bar{\rho}_{\lambda})/\bar{\rho}_{\lambda}$, we obtain a set of coupled integro-differential equations in the form of
\begin{equation}\label{Eq:integro-diff}
    \partial_\lambda \bar{\gamma}^{(n)}_{\lambda}
    = f^{(n)}  \left[  \bar{\gamma}^{(0)}_{\lambda}, \ldots , \bar{\gamma}^{(n)}_{\lambda}, \bar{\gamma}^{(n+1)}_{\lambda}, \bar{\gamma}^{(n+2)}_{\lambda} \right]\,,
\end{equation}
where $n=0, 1, 2, 3, \cdots$, and $\bar{\gamma}^{(1)}_{\lambda} \equiv 0$.
Note that $f^{(n)}$ depends not only on $\bar{\gamma}^{(0, \ldots,  n+2)}_{\lambda}$ but also on $\bar{\Gamma}^{(0, \ldots, n+2)}_{{\rm KS}, \lambda}$ and $\partial_{\lambda} \bar{\rho}_{\lambda}$ originating from the expansion.

A closed set of equations for $\bar{\gamma}^{(0, \ldots,  m)}_{\lambda}$ and $\bar{\rho}_{\lambda}$ is obtained from Eq.~(\ref{Eq:integro-diff}) under the $m$-th order truncation, $\bar{\gamma}^{(n \ge m+1)}_{\lambda}=0$.
In principle, the uncertainty of the $m$-th order solution can be checked by solving the $(m+1)$-th order equations.
However, it is not always possible to go higher orders in physical systems, so that a practical method of uncertainty quantification would be desirable.
Here we introduce a simple uncertainty estimate by taking only a first iteration of solving the $(m+1)$-th order equations.
First, we insert the $m$-th order results into the flow equation for $\bar{\gamma}^{(m+1)}_{\lambda}$.
Then we obtain an approximate solution, $\bar{\gamma}^{(m+1)}_{\lambda,\,{\rm app.}}$.
We plug this into the $m$-th order flow equations to obtain updated $m$-th order solutions.
The difference from the original ones is an uncertainty measure.
We will discuss an actual procedure to assign error bars to $\bar{\gamma}^{(0, \ldots, m)}_{\lambda}$ by using a simple model below.


\textit{0-D $\varphi^4$ theory.}---Let us now demonstrate how the KS-FRG works for obtaining energy functional in a simple 0-D bosonic model with the classical action,
\begin{equation}
  S[\varphi] = \frac{1}2 \omega^2\varphi^2 + \frac{1}{4!} v\varphi^4\,.
\end{equation}
Its generating function is just obtained by an ordinary integral
\begin{equation}
  e^{W[J]} = \sqrt{\frac{\omega^2}{2\pi}} \int_{-\infty}^{\infty} \mathrm{d} \varphi\, \exp\left(-S[\varphi]+J\varphi^2\right)\,.
\end{equation}
The exact solutions for the ground-state energy $E_{\rm gs}$ and the density $\rho_{\rm gs} = \langle \varphi^2 \rangle$ are known to be written in terms of the modified Bessel functions \cite{Kemler2013J.Phys.G40_085105}.

By taking $U_{{\rm KS},\lambda}=(\omega_{{\rm KS},\lambda})^2/2$ with $\omega_{{\rm KS}, \lambda=0}=\omega$, the mean-field part of the effective action becomes
\begin{equation}\label{eq:KS-0D}
  \Gamma_{{\rm KS},\lambda}[\rho] = \frac{1}{2}
  \left[ - {\ln (\omega^2\rho)} -1 +  (\omega_{{\rm KS},\lambda})^2  \rho \right]\,.
\end{equation}
In this case, Eq.~(\ref{Eq:KS-eq}) can be solved analytically to obtain $\bar{\rho}_{\lambda} = (1/\omega_{{\rm KS},{\lambda}} )^2$.
On the other hand, the flow equation~(\ref{Eq:FRG2}) for the correlation part reads
\begin{equation}\label{Eq:0D_FRG}
  \partial_\lambda   \gamma_{\lambda} [\rho]
  =  \rho  \bar{\Gamma}^{(2)}_{{\rm KS},\lambda } \partial_{\lambda} \bar{\rho}_{\lambda}
 + \frac{v}{4!}  [ \rho^2 + (   \Gamma^{(2)}_{{\rm KS},\lambda } [\rho] +  \gamma^{(2)}_{\lambda} [\rho]   )^{-1} ]\,.
\end{equation}
Combining Eqs.~(\ref{eq:KS-0D}) and (\ref{Eq:0D_FRG}), the ground-state energy of the system becomes
\begin{equation}
    E_{\rm gs}=
    \left[ -\frac{1}{2} {\ln (\omega^2\bar{\rho}_{\lambda}) }
    +  \bar{\gamma}_\lambda^{(0)}\right]_{\lambda=1}\,.
\end{equation}
Here $\bar{\rho}_{\lambda}$ and $\bar{\gamma}_\lambda^{(0)}$ are obtained by solving the flow equation~(\ref{Eq:0D_FRG}) up to a certain order.

For example, the equations up to $n=4$ are
\begin{widetext}
\begin{subequations}\label{Eq:KS4}
\begin{align}
  \partial_\lambda  \bar{\gamma}_\lambda^{(0)}
    &= \frac{v}{4!}\,\left( \bar{\rho}_{\lambda}^2 + \bar{G}_\lambda\right )+ \frac{1}{2\bar\rho_\lambda} (\partial_\lambda \bar{\rho}_{\lambda}) \,,\label{Eq:a}\\
  0
    &= \frac{v}{4!}\,\left( 2\bar{\rho}_{\lambda} \bar{G}_\lambda - \bar{\Gamma}_\lambda^{(3)} (\bar{G}_\lambda)^3\right)  + \partial_\lambda \bar{\rho}_{\lambda} \,,\label{Eq:b}\\
  \partial_\lambda  \bar{\gamma}_\lambda^{(2)}
    &= \frac{v}{4!} \left( 2 - \bar{\Gamma}_\lambda^{(4)}(\bar{G}_\lambda)^2  + 2 (\bar{\Gamma}_\lambda^{(3)})^2 (\bar{G}_\lambda)^3 \right) + \bar{\gamma}_\lambda^{(3)} (\partial_\lambda \bar{\rho}_{\lambda})\,,\label{Eq:c}\\
  \partial_\lambda  \bar{\gamma}_\lambda^{(3)}
    &= \frac{v}{4!}\left( - \bar{\Gamma}_\lambda^{(5)} (\bar{G}_\lambda)^2 + 6 \bar{\Gamma}_\lambda^{(3)} \bar{\Gamma}_\lambda^{(4)} (\bar{G}_\lambda)^3
    - 6 (\bar{\Gamma}_\lambda^{(3)})^3 (\bar{G}_\lambda)^4\right) + \bar{\gamma}_\lambda^{(4)} (\partial_\lambda \bar{\rho}_{\lambda})\,,\label{Eq:d}\\
  \partial_\lambda  \bar{\gamma}_\lambda^{(4)}
   &= \frac{v}{4!} \left( - \bar{\Gamma}_\lambda^{(6)}(\bar{G}_\lambda)^2 + (8 \bar{\Gamma}_\lambda^{(3)} \bar{\Gamma}_\lambda^{(5)} + 6 (\bar{\Gamma}_\lambda^{(4)})^2) (\bar{G}_\lambda)^3 - 36 (\bar{\Gamma}_\lambda^{(3)})^2 \bar{\Gamma}_\lambda^{(4)} (\bar{G}_\lambda)^4 + 24(\bar{\Gamma}_\lambda^{(3)})^4
   (\bar{G}_\lambda) ^5\right)
   + \bar{\gamma}_\lambda^{(5)} (\partial_\lambda \bar{\rho}_{\lambda})\,,\label{Eq:e}
\end{align}
\end{subequations}
\end{widetext}
where $\bar{\Gamma}^{(n)}_{\lambda }  = \bar{\Gamma}^{(n)}_{{\rm KS},\lambda } +  \bar{\gamma}^{(n)}_{\lambda}$  and  $\bar{G}_\lambda  \equiv (\bar{\Gamma}_\lambda^{(2)})^{-1}=(\bar{\Gamma}^{(2)}_{{\rm KS},\lambda } +  \bar{\gamma}^{(2)}_{\lambda})^{-1}$, with initial conditions $\bar{\rho}_{\lambda=0} =(1/ \omega)^{2}$ and $ \bar{\gamma}_{\lambda=0}^{(n)}= 0$.

Let us now discuss the uncertainty quantification by taking the third-order truncation as an example.
In this case, we first solve Eqs.~(\ref{Eq:a})--(\ref{Eq:d}) with $\bar{\gamma}_\lambda^{(n\ge 4)}=0$.
Then, the solutions $\bar{\gamma}_\lambda^{(0,\ldots, 3)}$ and $\bar{\rho}_{\lambda}$ together with $\bar{\gamma}_\lambda^{(5,6)}=0$  are used in the right-hand side of Eq.~(\ref{Eq:e}) to obtain an approximate solution $\bar{\gamma}_{\lambda,\,{\rm app.}}^{(4)}$.
Next we introduce an ansatz $g^{(4)}_\lambda \lambda$ that satisfies
\begin{equation}
  c_{_{\rm L}} g^{(4)}_\lambda \lambda \le  \bar{\gamma}_{\lambda,\,{\rm app.}}^{(4)} \le  c_{_{\rm U}} g^{(4)}_\lambda \lambda
\end{equation}
in the interval $\lambda \in [0,1]$.
Since we know $\bar{\gamma}_{\lambda=0}^{(4)}=0$, we separate out the factor $\lambda$ explicitly in the ansatz.
The constants $c_{_{\rm L,U}}$ are defined by
\begin{equation}
  c_{_{\rm L}} = \inf_{\lambda} \partial_\lambda(\bar{\gamma}_{\lambda,\,{\rm app.}}^{(4)}/g^{(4)}_\lambda)
  \quad\mbox{and}\quad
  c_{_{\rm U}} =  \sup_{\lambda} \partial_\lambda(\bar{\gamma}_{\lambda,\,{\rm app.}}^{(4)}/g^{(4)}_\lambda)\,.
\end{equation}
Natural choice of the ansatz  in the present model is $g^{(n)}_\lambda = v \bar{\rho}_{\lambda}^{2-n}$ obtained by inspecting the $v$ and $\bar{\rho}_{\lambda}$ dependence of the right-hand sides of Eqs.~(\ref{Eq:KS4}) and (\ref{eq:KS-0D}).
By substituting $c_{_{\rm L}} g^{(4)}_\lambda \lambda$, $c_{_{\rm U}} g^{(4)}_\lambda \lambda$, and $c_{_{\rm M}} g^{(4)}_\lambda \lambda$  with $c_{_{\rm M}} \equiv \bar{\gamma}_{1,\,{\rm app.}}^{(4)}/g^{(4)}_1$  for  $\bar{\gamma}_{\lambda}^{(4)} $ in Eqs.~(\ref{Eq:a})--(\ref{Eq:d}), we end up with most probable solutions for  $\bar{\gamma}_\lambda^{(0,\ldots, 3)}$ from $c_{_{\rm M}}$ and their errors from  $c_{_{\rm L,U}}$.
(In practice, we use $2c_{_{\rm L,U}}$ for conservative uncertainty estimates to take into account the effects from $\bar{\gamma}_\lambda^{(n\ge 5)}$, guided by the idea of effective field theory that the effects from higher orders should not be larger than those from the leading orders.)


\begin{figure}
  \centering
  \includegraphics[width=8cm]{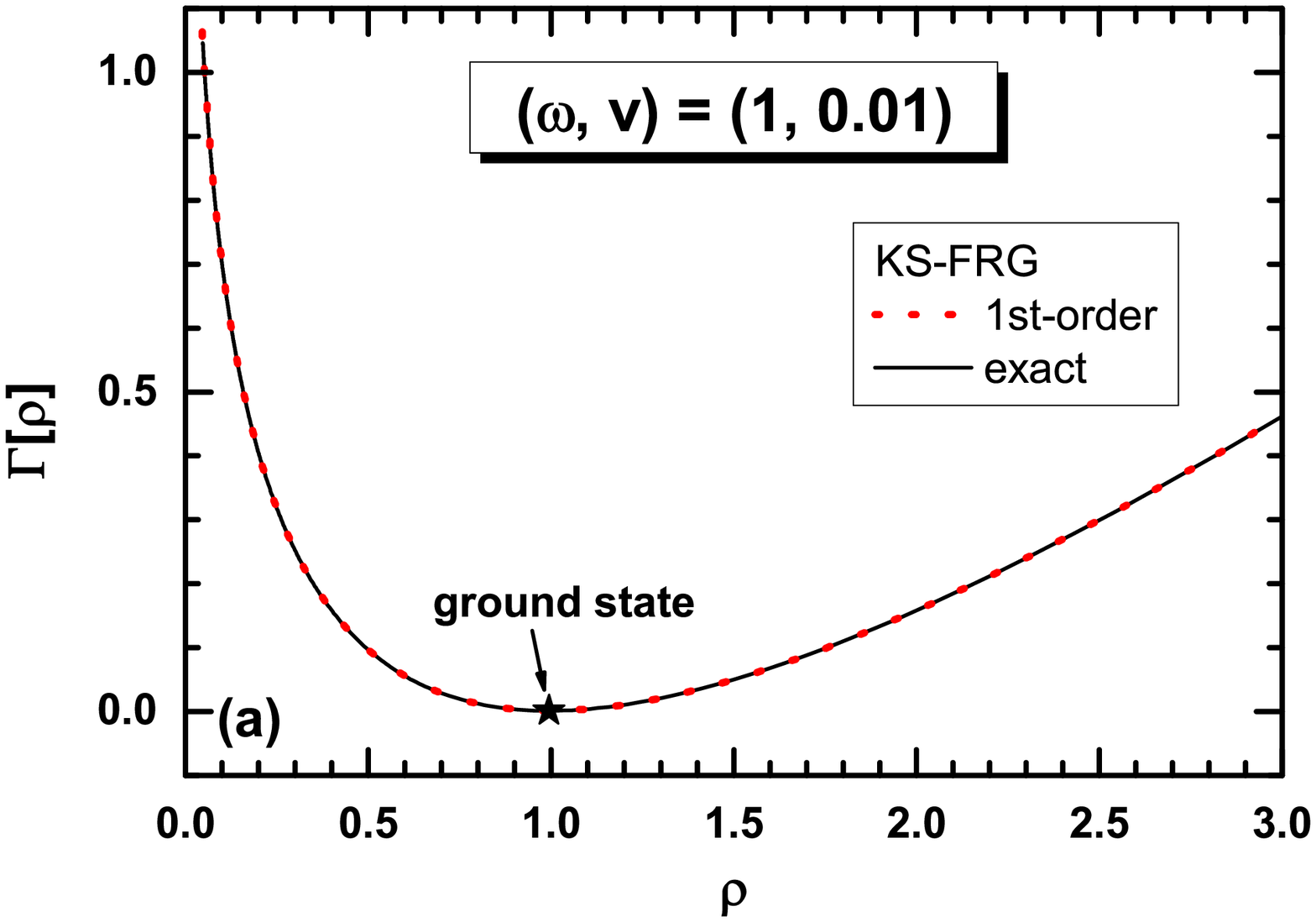}\\
  \includegraphics[width=8cm]{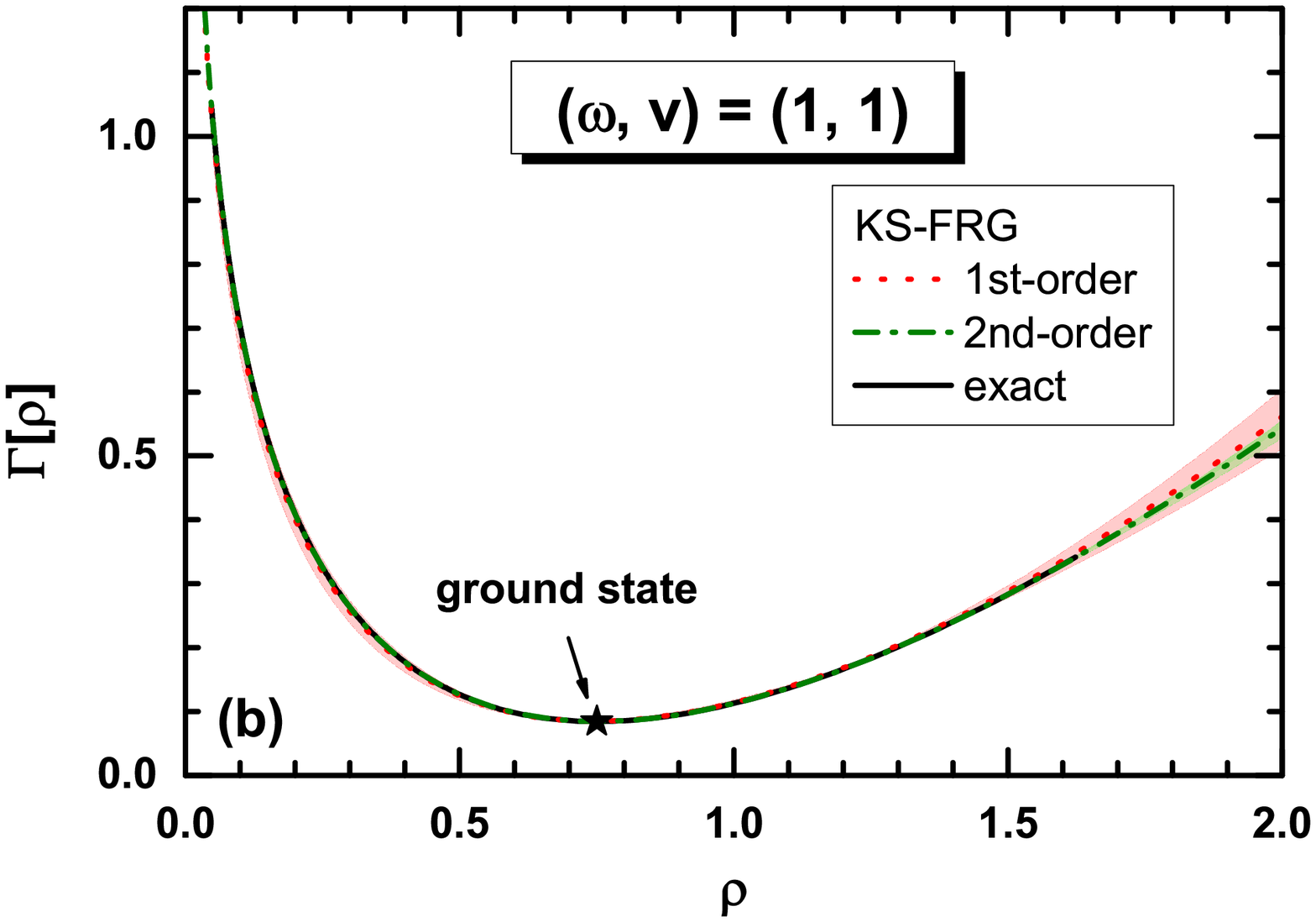}\\
  \includegraphics[width=8cm]{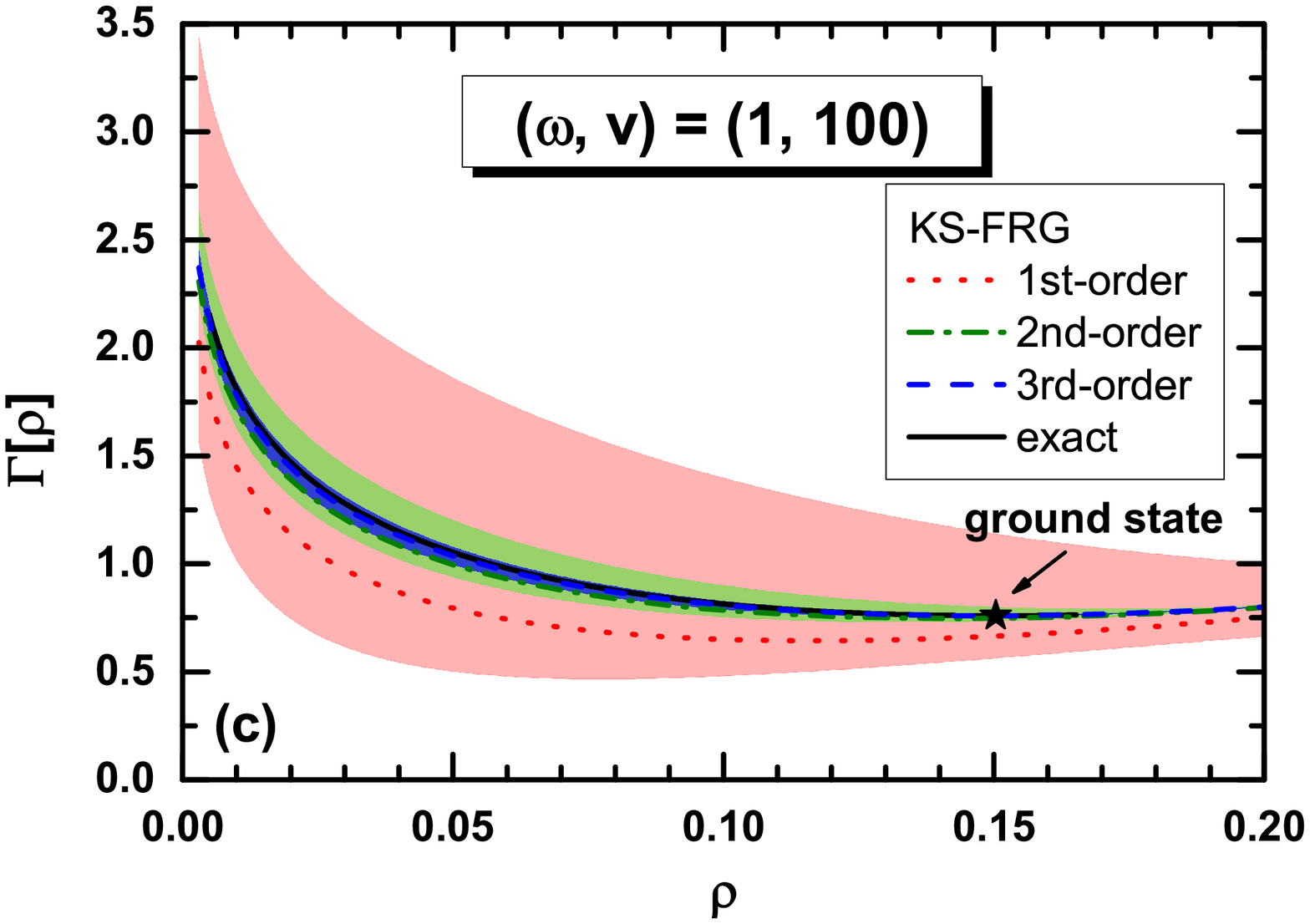}\\
  \caption{Effective actions as a function of density for (a) weak, (b) intermediate, and (c) strong couplings.
  The exact solutions are shown as the solid lines, while the 1st-, 2nd-, and 3rd-order KS-FRG results are shown as the red dotted, green dash-dotted, and blue dashed lines with the corresponding uncertainties as the shaded regions, respectively.}
  \label{Fig1}
\end{figure}

\textit{Numerical results.}---We take three typical cases: a weak coupling $(\omega,v)=(1,0.01)$, an intermediate coupling $(\omega,v)=(1,1)$, and a strong coupling $(\omega,v)=(1,100)$.
The last case has barely been discussed before in FRG.
The ground-state density $\rho_{\rm gs}=\bar{\rho}_{\lambda=1}$ and energy $E_{\rm gs}$ obtained by KS-FRG in the first-, second-, and third-order truncations are listed in Table~\ref{Table1} for the three cases.
Corresponding effective actions $\Gamma[\rho]$ as a function of $\rho$ are shown in Fig.~\ref{Fig1} with error bands at each order of truncation.
The exact solutions are shown by the solid lines for comparison.

In the weak-coupling case, the accuracy for $\rho_{\rm gs}$  and $E_{\rm gs}$ in the first-order calculation are already at $O(10^{-6})$ and $O(10^{-7})$ level, respectively, as shown in Table~\ref{Table1}.
Also, $\Gamma[\rho]$ in the first order is already on top of the exact solution in a very wide density range with invisible theoretical uncertainty as shown in Fig.~\ref{Fig1}(a).

In the intermediate-coupling case, an order of magnitude improvement of the accuracy of $\rho_{\rm gs}$ and $E_{\rm gs}$ is seen by increasing the order of truncation.
The third-order calculation of $E_{\rm gs}$ reaches to $O(10^{-4})$ accuracy in KS-FRG as shown in Table~\ref{Table1}.
This is in contrast to the conventional FRG calculation which gives only $O(10^{-2})$ accuracy even with a 6th-order calculation \cite{Kemler2013J.Phys.G40_085105}.
The rapid convergence and the rapid shrinking of the error in KS-FRG are also found for $\Gamma[\rho]$ as shown in Fig.~\ref{Fig1}(b).

Even in the strong-coupling case, an order of magnitude improvement of the accuracy is achieved by increasing the order of truncation.
The third-order results of $\rho_{\rm gs}$ and $E_{\rm gs}$ reach to $O(10^{-2})$ accuracy as shown in Table~\ref{Table1}.
The convergence of $\Gamma[\rho]$ to the exact result is also seen clearly in Fig.~\ref{Fig1}(c).

\begin{figure}
  \centering
  \includegraphics[width=8cm]{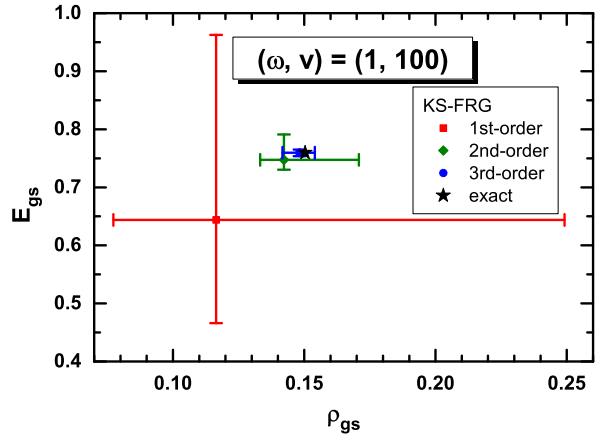}
  \caption{Ground-state energy versus ground-state density for the strong-coupling case.
  The results by the 1st-, 2nd-, and 3rd-order KS-FRG calculations are shown as the square, diamond, and circle with the theoretical uncertainties, while the exact value is shown with the star.}
  \label{Fig2}
\end{figure}

Such a rapid convergence in our KS-FRG scheme, as also illustrated in Fig.~\ref{Fig2} for the strong-coupling case, stems from the facts that significant part of ${\Gamma}_{\lambda}[\rho]$ is already taken into account in the mean-field part ${\Gamma}_{{\rm KS},\lambda}[\rho]$ which evolves with $\lambda$, and the correlation part ${\gamma}_{{\rm KS},\lambda}[\rho]$ can be treated well as small fluctuations around the mean-field part.

\begin{figure}
  \centering
  \includegraphics[width=8cm]{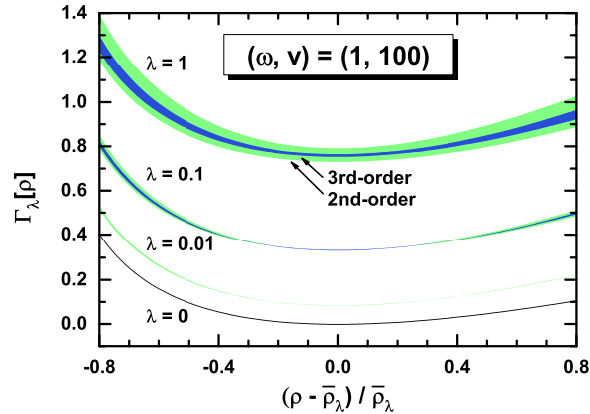}
  \caption{The effective action $\Gamma_\lambda[\rho]$ with $\lambda=0$, $0.01$, $0.1$, and $1$ as a function of $(\rho-\bar{\rho}_{\lambda})/\bar{\rho}_{\lambda}$ for the strong-coupling case.
  The 2nd- and 3rd-order KS-FRG results with uncertainties are shown by the green and blue bands, respectively. }
  \label{Fig3}
\end{figure}

In Fig.~\ref{Fig3}, we show how the effective action $\Gamma_\lambda[\rho]$ and its uncertainty in the strong-coupling case evolve under the FRG flow from the non-interacting system at $\lambda=0$ to the fully interacting system at $\lambda=1$.
Due to the repulsive nature of the interaction, the effective action increases as $\lambda$ increases.
Also, the uncertainty grows as $\lambda$ increases because of the truncation of the coupled flow equations.
From Eq.~(\ref{Eq:integro-diff}) it is seen that the uncertainties in $\bar\gamma_\lambda^{(n+1)}$ and $\bar\gamma_\lambda^{(n+2)}$ propagate to $\bar\gamma_\lambda^{(n)}$ with the flow evolution.
In such a way, the total truncation uncertainties in $\Gamma_\lambda[\rho]$ propagate from high- and low-density regions towards the stationary point $\bar\rho_{\lambda}$, controlled by the power counting $[(\rho-\bar{\rho}_{\lambda})/\bar{\rho}_{\lambda}]^n$.
The speed of propagation depends on the strength of interaction.


\textit{Summary.}---In this Letter, we have proposed a novel optimization method of FRG in analogy with the Kohn-Sham scheme of DFT.
Essential idea of our method, called KS-FRG, is to separate the full effective action into the mean-field (KS) part and the correlation part at each flow parameter $\lambda$.
Then the KS equation for the mean field and the FRG flow equation for the correlation are solved self-consistently.
In practice, the correlation part is expanded in Tayler series around the stationary point $\bar{\rho}_{\lambda}$ of the KS effective action.
This leads to rapid convergence of the energy density functional in comparison to the conventional FRG method.
Furthermore, we have presented a simple method to estimate the truncation uncertainty of the coupled flow equation.
By taking the $\varphi^4$ theory in 0-D as a benchmark, we demonstrated explicitly that this KS-FRG provides not only high-precision calculation of the energy density functional but also a useful uncertainty measure for the results.
This method is a promising candidate for making systematic and fast converging calculations of the quantum many-body systems, such as the cold atoms near unitarity, nucleons in finite nuclei, and so on.
One of our next steps is to apply the method to models with temporal and spatial degrees of freedom, which will be  reported elsewhere.

We are grateful to the stimulating discussions with Wolfram Weise.
This work is partly supported by the RIKEN iTHES and iTHEMS Programs.
T.H. is grateful to the Aspen Center for Physics, supported in part by NSF Grants PHY1066292 and PHY1607611.


\begin{thebibliography}{25}
\expandafter\ifx\csname natexlab\endcsname\relax\def\natexlab#1{#1}\fi
\expandafter\ifx\csname bibnamefont\endcsname\relax
  \def\bibnamefont#1{#1}\fi
\expandafter\ifx\csname bibfnamefont\endcsname\relax
  \def\bibfnamefont#1{#1}\fi
\expandafter\ifx\csname citenamefont\endcsname\relax
  \def\citenamefont#1{#1}\fi
\expandafter\ifx\csname url\endcsname\relax
  \def\url#1{\texttt{#1}}\fi
\expandafter\ifx\csname urlprefix\endcsname\relax\def\urlprefix{URL }\fi
\providecommand{\bibinfo}[2]{#2}
\providecommand{\eprint}[2][]{\url{#2}}

\bibitem[{\citenamefont{Hohenberg and
  Kohn}(1964)}]{Hohenberg1964Phys.Rev.136_B864--B871}
\bibinfo{author}{\bibfnamefont{P.}~\bibnamefont{Hohenberg}} \bibnamefont{and}
  \bibinfo{author}{\bibfnamefont{W.}~\bibnamefont{Kohn}},
  \bibinfo{journal}{Phys. Rev.} \textbf{\bibinfo{volume}{136}},
  \bibinfo{pages}{B864} (\bibinfo{year}{1964}).

\bibitem[{\citenamefont{Kohn and
  Sham}(1965)}]{Kohn1965Phys.Rev.140_A1133--A1138}
\bibinfo{author}{\bibfnamefont{W.}~\bibnamefont{Kohn}} \bibnamefont{and}
  \bibinfo{author}{\bibfnamefont{L.~J.} \bibnamefont{Sham}},
  \bibinfo{journal}{Phys. Rev.} \textbf{\bibinfo{volume}{140}},
  \bibinfo{pages}{A1133} (\bibinfo{year}{1965}).

\bibitem[{\citenamefont{Kohn}(1999)}]{Kohn1999Rev.Mod.Phys.71_1253--1266}
\bibinfo{author}{\bibfnamefont{W.}~\bibnamefont{Kohn}}, \bibinfo{journal}{Rev.
  Mod. Phys.} \textbf{\bibinfo{volume}{71}}, \bibinfo{pages}{1253}
  (\bibinfo{year}{1999}).

\bibitem[{\citenamefont{Kryachko and
  Lude\~{n}a}(2014)}]{Kryachko2014Phys.Rep.544_123--239}
\bibinfo{author}{\bibfnamefont{E.~S.} \bibnamefont{Kryachko}} \bibnamefont{and}
  \bibinfo{author}{\bibfnamefont{E.~V.} \bibnamefont{Lude\~{n}a}},
  \bibinfo{journal}{Phys. Rep.} \textbf{\bibinfo{volume}{544}},
  \bibinfo{pages}{123} (\bibinfo{year}{2014}).

\bibitem[{\citenamefont{Zangwill}(2015)}]{Zangwill2015Phys.Today68_34--39}
\bibinfo{author}{\bibfnamefont{A.}~\bibnamefont{Zangwill}},
  \bibinfo{journal}{Phys. Today} \textbf{\bibinfo{volume}{68}},
  \bibinfo{pages}{34} (\bibinfo{year}{2015}).

\bibitem[{\citenamefont{Jones}(2015)}]{Jones2015Rev.Mod.Phys.87_897--923}
\bibinfo{author}{\bibfnamefont{R.~O.} \bibnamefont{Jones}},
  \bibinfo{journal}{Rev. Mod. Phys.} \textbf{\bibinfo{volume}{87}},
  \bibinfo{pages}{897} (\bibinfo{year}{2015}).

\bibitem[{\citenamefont{Kutzelnigg}(2006)}]{Kutzelnigg2006J.Mol.Struct.768_163--173}
\bibinfo{author}{\bibfnamefont{W.}~\bibnamefont{Kutzelnigg}},
  \bibinfo{journal}{J. Mol. Struct.} \textbf{\bibinfo{volume}{768}},
  \bibinfo{pages}{163} (\bibinfo{year}{2006}).

\bibitem[{\citenamefont{Berland et~al.}(2015)\citenamefont{Berland, Cooper,
  Lee, Schr\"oder, Thonhauser, Hyldgaard, and
  Lundqvist}}]{Berland2015Rep.Prog.Phys.78_066501}
\bibinfo{author}{\bibfnamefont{K.}~\bibnamefont{Berland}},
  \bibinfo{author}{\bibfnamefont{V.~R.} \bibnamefont{Cooper}},
  \bibinfo{author}{\bibfnamefont{K.}~\bibnamefont{Lee}},
  \bibinfo{author}{\bibfnamefont{E.}~\bibnamefont{Schr\"oder}},
  \bibinfo{author}{\bibfnamefont{T.}~\bibnamefont{Thonhauser}},
  \bibinfo{author}{\bibfnamefont{P.}~\bibnamefont{Hyldgaard}},
  \bibnamefont{and} \bibinfo{author}{\bibfnamefont{B.~I.}
  \bibnamefont{Lundqvist}}, \bibinfo{journal}{Rep. Prog. Phys.}
  \textbf{\bibinfo{volume}{78}}, \bibinfo{pages}{066501}
  (\bibinfo{year}{2015}).

\bibitem[{\citenamefont{Drut et~al.}(2010)\citenamefont{Drut, Furnstahl, and
  Platter}}]{Drut2010Prog.Part.Nucl.Phys.64_120--168}
\bibinfo{author}{\bibfnamefont{J.~E.} \bibnamefont{Drut}},
  \bibinfo{author}{\bibfnamefont{R.~J.} \bibnamefont{Furnstahl}},
  \bibnamefont{and} \bibinfo{author}{\bibfnamefont{L.}~\bibnamefont{Platter}},
  \bibinfo{journal}{Prog. Part. Nucl. Phys.} \textbf{\bibinfo{volume}{64}},
  \bibinfo{pages}{120} (\bibinfo{year}{2010}).

\bibitem[{\citenamefont{Nakatsukasa et~al.}(2016)\citenamefont{Nakatsukasa,
  Matsuyanagi, Matsuo, and Yabana}}]{Nakatsukasa2016Rev.Mod.Phys.88_045004}
\bibinfo{author}{\bibfnamefont{T.}~\bibnamefont{Nakatsukasa}},
  \bibinfo{author}{\bibfnamefont{K.}~\bibnamefont{Matsuyanagi}},
  \bibinfo{author}{\bibfnamefont{M.}~\bibnamefont{Matsuo}}, \bibnamefont{and}
  \bibinfo{author}{\bibfnamefont{K.}~\bibnamefont{Yabana}},
  \bibinfo{journal}{Rev. Mod. Phys.} \textbf{\bibinfo{volume}{88}},
  \bibinfo{pages}{045004} (\bibinfo{year}{2016}).

\bibitem[{\citenamefont{Cohen et~al.}(2012)\citenamefont{Cohen, Mori-S\'anchez,
  and Yang}}]{Cohen2012Chem.Rev.112_289--320}
\bibinfo{author}{\bibfnamefont{A.~J.} \bibnamefont{Cohen}},
  \bibinfo{author}{\bibfnamefont{P.}~\bibnamefont{Mori-S\'anchez}},
  \bibnamefont{and} \bibinfo{author}{\bibfnamefont{W.}~\bibnamefont{Yang}},
  \bibinfo{journal}{Chem. Rev.} \textbf{\bibinfo{volume}{112}},
  \bibinfo{pages}{289} (\bibinfo{year}{2012}).

\bibitem[{\citenamefont{Himmetoglu et~al.}(2014)\citenamefont{Himmetoglu,
  Floris, de~Gironcoli, and
  Cococcioni}}]{Himmetoglu2014Int.J.QuantumChem.114_14--49}
\bibinfo{author}{\bibfnamefont{B.}~\bibnamefont{Himmetoglu}},
  \bibinfo{author}{\bibfnamefont{A.}~\bibnamefont{Floris}},
  \bibinfo{author}{\bibfnamefont{S.}~\bibnamefont{de~Gironcoli}},
  \bibnamefont{and}
  \bibinfo{author}{\bibfnamefont{M.}~\bibnamefont{Cococcioni}},
  \bibinfo{journal}{Int. J. Quantum Chem.} \textbf{\bibinfo{volume}{114}},
  \bibinfo{pages}{14} (\bibinfo{year}{2014}).

\bibitem[{\citenamefont{Sperger et~al.}(2015)\citenamefont{Sperger, Sanhueza,
  Kalvet, and Schoenebeck}}]{Sperger2015Chem.Rev.115_9532--9586}
\bibinfo{author}{\bibfnamefont{T.}~\bibnamefont{Sperger}},
  \bibinfo{author}{\bibfnamefont{I.~A.} \bibnamefont{Sanhueza}},
  \bibinfo{author}{\bibfnamefont{I.}~\bibnamefont{Kalvet}}, \bibnamefont{and}
  \bibinfo{author}{\bibfnamefont{F.}~\bibnamefont{Schoenebeck}},
  \bibinfo{journal}{Chem. Rev.} \textbf{\bibinfo{volume}{115}},
  \bibinfo{pages}{9532} (\bibinfo{year}{2015}).

\bibitem[{\citenamefont{Erler et~al.}(2012)\citenamefont{Erler, Birge,
  Kortelainen, Nazarewicz, Olsen, Perhac, and
  Stoitsov}}]{Erler2012Nature486_509--512}
\bibinfo{author}{\bibfnamefont{J.}~\bibnamefont{Erler}},
  \bibinfo{author}{\bibfnamefont{N.}~\bibnamefont{Birge}},
  \bibinfo{author}{\bibfnamefont{M.}~\bibnamefont{Kortelainen}},
  \bibinfo{author}{\bibfnamefont{W.}~\bibnamefont{Nazarewicz}},
  \bibinfo{author}{\bibfnamefont{E.}~\bibnamefont{Olsen}},
  \bibinfo{author}{\bibfnamefont{A.~M.} \bibnamefont{Perhac}},
  \bibnamefont{and} \bibinfo{author}{\bibfnamefont{M.}~\bibnamefont{Stoitsov}},
  \bibinfo{journal}{Nature} \textbf{\bibinfo{volume}{486}},
  \bibinfo{pages}{509} (\bibinfo{year}{2012}).

\bibitem[{\citenamefont{Dobaczewski et~al.}(2014)\citenamefont{Dobaczewski,
  Nazarewicz, and Reinhard}}]{Dobaczewski2014J.Phys.G41_074001}
\bibinfo{author}{\bibfnamefont{J.}~\bibnamefont{Dobaczewski}},
  \bibinfo{author}{\bibfnamefont{W.}~\bibnamefont{Nazarewicz}},
  \bibnamefont{and} \bibinfo{author}{\bibfnamefont{P.-G.}
  \bibnamefont{Reinhard}}, \bibinfo{journal}{J. Phys. G}
  \textbf{\bibinfo{volume}{41}}, \bibinfo{pages}{074001}
  (\bibinfo{year}{2014}).

\bibitem[{\citenamefont{McDonnell et~al.}(2015)\citenamefont{McDonnell,
  Schunck, Higdon, Sarich, Wild, and
  Nazarewicz}}]{McDonnell2015Phys.Rev.Lett.114_122501}
\bibinfo{author}{\bibfnamefont{J.~D.} \bibnamefont{McDonnell}},
  \bibinfo{author}{\bibfnamefont{N.}~\bibnamefont{Schunck}},
  \bibinfo{author}{\bibfnamefont{D.}~\bibnamefont{Higdon}},
  \bibinfo{author}{\bibfnamefont{J.}~\bibnamefont{Sarich}},
  \bibinfo{author}{\bibfnamefont{S.~M.} \bibnamefont{Wild}}, \bibnamefont{and}
  \bibinfo{author}{\bibfnamefont{W.}~\bibnamefont{Nazarewicz}},
  \bibinfo{journal}{Phys. Rev. Lett.} \textbf{\bibinfo{volume}{114}},
  \bibinfo{pages}{122501} (\bibinfo{year}{2015}).

\bibitem[{\citenamefont{Nazarewicz}(2016)}]{Nazarewicz2016J.Phys.G43_044002}
\bibinfo{author}{\bibfnamefont{W.}~\bibnamefont{Nazarewicz}},
  \bibinfo{journal}{J. Phys. G} \textbf{\bibinfo{volume}{43}},
  \bibinfo{pages}{044002} (\bibinfo{year}{2016}).

\bibitem[{\citenamefont{Wetterich}(1993)}]{Wetterich1993Phys.Lett.B301_90--94}
\bibinfo{author}{\bibfnamefont{C.}~\bibnamefont{Wetterich}},
  \bibinfo{journal}{Phys. Lett. B} \textbf{\bibinfo{volume}{301}},
  \bibinfo{pages}{90} (\bibinfo{year}{1993}).

\bibitem[{\citenamefont{Metzner et~al.}(2012)\citenamefont{Metzner, Salmhofer,
  Honerkamp, Meden, and Sch\"onhammer}}]{Metzner2012Rev.Mod.Phys.84_299--352}
\bibinfo{author}{\bibfnamefont{W.}~\bibnamefont{Metzner}},
  \bibinfo{author}{\bibfnamefont{M.}~\bibnamefont{Salmhofer}},
  \bibinfo{author}{\bibfnamefont{C.}~\bibnamefont{Honerkamp}},
  \bibinfo{author}{\bibfnamefont{V.}~\bibnamefont{Meden}}, \bibnamefont{and}
  \bibinfo{author}{\bibfnamefont{K.}~\bibnamefont{Sch\"onhammer}},
  \bibinfo{journal}{Rev. Mod. Phys.} \textbf{\bibinfo{volume}{84}},
  \bibinfo{pages}{299} (\bibinfo{year}{2012}).

\bibitem[{\citenamefont{Polonyi and
  Sailer}(2002)}]{Polonyi2002Phys.Rev.B66_155113}
\bibinfo{author}{\bibfnamefont{J.}~\bibnamefont{Polonyi}} \bibnamefont{and}
  \bibinfo{author}{\bibfnamefont{K.}~\bibnamefont{Sailer}},
  \bibinfo{journal}{Phys. Rev. B} \textbf{\bibinfo{volume}{66}},
  \bibinfo{pages}{155113} (\bibinfo{year}{2002}).

\bibitem[{\citenamefont{Schwenk and Polonyi}(2004)}]{Schwenk2004_273-282}
\bibinfo{author}{\bibfnamefont{A.}~\bibnamefont{Schwenk}} \bibnamefont{and}
  \bibinfo{author}{\bibfnamefont{J.}~\bibnamefont{Polonyi}}, in
  \emph{\bibinfo{booktitle}{{32nd International Workshop on Gross Properties of
  Nuclei and Nuclear Excitation: Probing Nuclei and Nucleons with Electrons and
  Photons}}} (\bibinfo{year}{2004}), pp. \bibinfo{pages}{273--282},
  \bibinfo{note}{arXiv:0403011 (nucl-th)}.

\bibitem[{\citenamefont{Braun}(2012)}]{Braun2012J.Phys.G39_033001}
\bibinfo{author}{\bibfnamefont{J.}~\bibnamefont{Braun}}, \bibinfo{journal}{J.
  Phys. G} \textbf{\bibinfo{volume}{39}}, \bibinfo{pages}{033001}
  (\bibinfo{year}{2012}).

\bibitem[{\citenamefont{Kemler and Braun}(2013)}]{Kemler2013J.Phys.G40_085105}
\bibinfo{author}{\bibfnamefont{S.}~\bibnamefont{Kemler}} \bibnamefont{and}
  \bibinfo{author}{\bibfnamefont{J.}~\bibnamefont{Braun}}, \bibinfo{journal}{J.
  Phys. G} \textbf{\bibinfo{volume}{40}}, \bibinfo{pages}{085105}
  (\bibinfo{year}{2013}).

\bibitem[{\citenamefont{Kemler et~al.}(2017)\citenamefont{Kemler, Pospiech, and
  Braun}}]{Kemler2017J.Phys.G44_015101}
\bibinfo{author}{\bibfnamefont{S.}~\bibnamefont{Kemler}},
  \bibinfo{author}{\bibfnamefont{M.}~\bibnamefont{Pospiech}}, \bibnamefont{and}
  \bibinfo{author}{\bibfnamefont{J.}~\bibnamefont{Braun}}, \bibinfo{journal}{J.
  Phys. G} \textbf{\bibinfo{volume}{44}}, \bibinfo{pages}{015101}
  (\bibinfo{year}{2017}).

\bibitem[{\citenamefont{Rentrop et~al.}(2015)\citenamefont{Rentrop, Jakobs, and
  Meden}}]{Rentrop2015J.Phys.A48_145002}
\bibinfo{author}{\bibfnamefont{J.~F.} \bibnamefont{Rentrop}},
  \bibinfo{author}{\bibfnamefont{S.~G.} \bibnamefont{Jakobs}},
  \bibnamefont{and} \bibinfo{author}{\bibfnamefont{V.}~\bibnamefont{Meden}},
  \bibinfo{journal}{J. Phys. A} \textbf{\bibinfo{volume}{48}},
  \bibinfo{pages}{145002} (\bibinfo{year}{2015}).

\end{thebibliography}

\end{document}